\documentstyle[aps,epsf,times]{revtex}
\newcommand{\dd}{\mbox{d}}
\newcommand{\DD}{\mbox{D}}
\newcommand{\tfrac}[2]{{\textstyle\frac{#1}{#2}}}
\begin{document} 
\title{Time-Dependent Ginzburg-Landau Theory and \protect\\
Duality} 
\author{Adriaan M. J. Schakel \\
Institut f\"ur Theoretische Physik \\ Freie Universit\"at
Berlin \\ Arnimallee 14, 14195 Berlin}  
\maketitle
\begin{abstract}
In the first part of this review paper, the time-dependent
Ginzburg-Landau theory is derived starting from the microscopic BCS
model with the help of a derivative expansion.  Special attention is
paid to two space dimensions, where the entire crossover from the
weak-coupling BCS limit to the strong-coupling BEC limit of tightly
bound fermion pairs is accessible analytically.  The second part deals
with the dual approach to the time-independent Ginzburg-Landau theory in
three space dimensions.  In this approach, the magnetic vortices of a
superconductor play the central role, and the superconductor-to-normal
phase transition is understood as a proliferation of these vortices.
\end{abstract}
%
\date{\today}
\section{Introduction} \label{intro}
The role of topological defects in symmetry-breaking phase transitions
cannot be underestimated.  The prime example of a defect-driven
(equilibrium) transition is the superfluid-to-normal phase transition in
a $^4$He film, which can be understood as the unbinding of
vortex-antivortex pairs \cite{Berezinskii,KT73}.  Since the very
existence of an ordered state and the presence of topological defects
are both manifestations of spontaneously broken symmetries, it doesn't
come as a surprise that various phase transitions can be understand in
terms of defects.  To appreciate this point of view, note that defects
are regions, usually macroscopic in size, where the symmetry is broken
differently from that in the rest of the system.  Often, the symmetry is
completely restored in the defect cores, meaning that they are in the
normal state.  (In systems like superfluid $^3$He with a large symmetry
group, the defect cores may be ordered states themselves, but of a
different character.)  Now, when defects with a normal-state core
proliferate, the ordered state gets disordered and converted to the
normal state.  

In recent years also the dynamics of phase transitions and the
accompanying formation of defects have received much attention (see
Ref.\ \cite{nato} for reviews).  An important role is played here by the
so-called time-dependent Ginzburg-Landau theory, which provides a
phenomenological approach.  Ideally, one would like to start with a
microscopic model of the system under consideration and derive the
time-dependent Ginzburg-Landau theory by integrating out irrelevant
degrees of freedom, thus turning the phenomenological theory into an
effective one.  For a superconductor this program, at least in the
time-independent case, has been carried out by Gorkov \cite{Gorkov}.
Starting from the microscopic theory of Bardeen, Cooper, and Schrieffer
(BCS) \cite{BCS}, he derived by means of a Green function method the
Ginzburg-Landau theory of superconductivity which had been proposed as a
phenomenological theory seven years before the BCS theory was
formulated \cite{GL}.  

In the first part of this review paper, we reformulate Gorkov's
derivation (extended to include time dependence) in terms of the
functional-integral approach to quantum field theory.  We will not only
study the weak-coupling BCS limit of loosely bound Cooper pairs, but
consider arbitrary values of the coupling constant $\lambda_0$,
including the limit $\lambda_0 \rightarrow -\infty$, where the fermions
form tightly bound pairs.  When going from the weak-coupling to the
large-$\lambda_0$ limit, the dynamics changes from dissipative to
non-dissipative behavior.

In the second part of this review paper, we discuss the dual approach to the
time-independent Ginzburg-Landau theory.  It is a formulation directly in
terms of magnetic vortices, in which the superconductor-to-normal is
described as a proliferation of vortices.

\subsection{Notation}
We adopt Feynman's notation and denote a spacetime point by $x=x_\mu
=(t,{\bf x})$, $\mu = 0,1, \cdots,d$, with $d$ the number of space
dimensions, while the energy $k_0$ and momentum ${\bf k}$ will be denoted by
$k=k_\mu = (k_0,{\bf k})$.  The time derivative $\partial_0 =
\partial/\partial t$ and the gradient $\nabla$ are sometimes combined in a
single vector $\tilde{\partial}_\mu = (\partial_0, -\nabla)$.  The tilde on
$\partial_\mu$ is to alert the reader for the minus sign appearing in the
spatial components of this vector.  We define the scalar product $k \cdot x
= k_\mu x_\mu = k_\mu g_{\mu \nu} k_\nu = k_0 t - {\bf k} \cdot {\bf x}$,
with $g_{\mu \nu} = {\rm diag}(1,-1, \cdots,-1)$ and use Einstein's
summation convention.  Because of the minus sign in the definition of the
vector $\tilde{\partial}_\mu$, it follows that $\tilde{\partial}_\mu a_\mu =
\partial_0 a_0 + \nabla \cdot {\bf a}$, with $a_\mu$ an arbitrary vector.

Integrals over spacetime are denoted by
$$
\int_{x} = \int_{t,{\bf x}} = \int \dd t \, \dd^d x,
$$
while those over energy and momentum by
$$
\int_k = \int_{k_0,{\bf k}} = \int \frac{\dd k_0}{2 \pi}
\frac{\dd^d k}{(2 \pi)^d}.
$$
When no integration limits are indicated, the integrals are assumed to
be over all possible values of the integration variables.

Natural units $\hbar = k_{\rm B} = 1$ are adopted throughout.

\section{Time-Dependent Ginzburg-Landau Theory} \label{sec:TDGL}
In this section, we derive the time-dependent Ginzburg-Landau theory for a
superconductor, starting from the microscopic BCS model.  Following Ref.\
\cite{thesis}, we integrate out the fermionic degrees of freedom in favor of
a bosonic field describing the fermion pairs and use a derivative expansion
method.  We wish to obtain the Ginzburg-Landau theory not only in the
weak-coupling BCS limit of loosely bound and overlapping Cooper pairs, but
for arbitrary values of the coupling constant, in particular in the limit
$\lambda_0 \rightarrow - \infty$ where the fermions form tightly bound pairs
\cite{Eagles,Leggett}.  This crossover between the two limits has recently
been studied in detail \cite{DrZw,Haussmann,MRE,MPS}.  Since in two space
dimensions this region becomes accessible by analytical methods, we shall,
unless stated otherwise, restrict ourselves to $d=2$ when carrying out
momentum integrals.

\subsection{BCS Model \label{sec:bcs}}
The BCS Lagrangian reads \cite{Popov}
\begin{eqnarray}  \label{bcs:BCS}
     {\cal L} &=& \psi^{\ast}_{\uparrow} [i\partial_0 - \xi(-i \nabla)]
\psi_{\uparrow}  
     + \psi_{\downarrow}^{\ast} [i \partial_0 - \xi(-i \nabla)]\psi_{\downarrow}
     - \lambda_0 \psi_{\uparrow}^{\ast}\,\psi_{\downarrow}
     ^{\ast}\,\psi_{\downarrow}\,\psi_{\uparrow}          \nonumber  \\
     &=:& {\cal L}_{0} + {\cal L}_{\rm i},                  
\end{eqnarray} 
where ${\cal L}_{\rm i} = - \lambda_0 \psi_{\uparrow}^{\ast} \,
\psi_{\downarrow}^{\ast}\,\psi_{\downarrow}\,\psi_{\uparrow}$ is a
contact interaction term, representing the effective attraction between
electrons with bare coupling constant $\lambda_0 < 0$, and ${\cal
L}_{0}$ is the remainder.  In Eq.\ (\ref{bcs:BCS}), the field $\psi_{\uparrow
(\downarrow )}$ is an anticommuting field describing the electrons with
mass $m$ and spin up (down); $\xi(-i \nabla) = \epsilon(-i \nabla) -
\mu_0$, with $\epsilon(-i \nabla) = - \nabla^2/2m$, is the kinetic
energy operator with the bare chemical potential $\mu_0$ subtracted.

For computational convenience we introduce Nambu's notation and rewrite the
Lagrangian (\ref{bcs:BCS}) in terms of a two-component field
\begin{equation} \label{32}
\psi = \left( \begin{array}{c} \psi_{\uparrow} \\ 
           \psi_{\downarrow}^{\ast}  \end{array} \right) \:\:\:\:\:\:
    \psi^{\dagger} = (\psi_{\uparrow}^{\ast},\psi_{\downarrow}).
\end{equation} 
In order to integrate out the fermionic degrees of freedom, the
zero-temperature partition function represented as a functional integral
\begin{equation}     \label{bcs:34}
Z = \int {\rm D} \psi^{\dagger} {\rm D} \psi \exp \left( i \int_x
\,{\cal L} \right),                                            
\end{equation} 
must be written in a form bilinear in the electron fields.  This is achieved
by rewriting the quartic interaction term as a functional integral over
auxiliary fields $\Delta$ and $\Delta^*$:
\begin{eqnarray}   \label{bcs:35} 
\lefteqn{
\exp \left( -i \lambda_0 \int_x \psi_{\uparrow}^{\ast}
\, \psi_{\downarrow}^{\ast} \, \psi_{\downarrow}\, \psi_{\uparrow} 
\right)  = }                                           \\
& &  \int {\rm D} \Delta^* {\rm D} \Delta \exp \left[ -i
\int_x \left( \Delta^* \, \psi_{\downarrow}\,\psi_{\uparrow} +
\psi_{\uparrow}^{\ast} \, \psi_{\downarrow}^{\ast} \, \Delta -
\frac{1}{\lambda_0 } |\Delta|^2 \right) \right]. \nonumber 
\end{eqnarray} 
The field equation for $\Delta^*$,
\begin{equation}  \label{bcs:del}
\Delta = \lambda_0 \psi_{\downarrow} \psi_{\uparrow},       
\end{equation} 
shows that the auxiliary field describes electron pairs.  We will therefore
refer to it as pair field.  The partition function thus becomes
\begin{eqnarray}  \label{bcs:36}
\lefteqn{
Z = \int {\rm D} \psi^{\dagger} {\rm D} \psi \int  {\rm D} \Delta^* {\rm
D} \Delta \; \exp\left(\frac{i}{\lambda_0} \int_x |\Delta|^2 \right)} \\
& & \times \exp \left[ i \int_x \, \psi^{\dagger} \left( \begin{array}{cc} i
\partial_{0} - \xi(-i \nabla) & -\Delta \\ -\Delta^* & i \partial_{0} +
\xi(-i \nabla)
\end{array} \right)   \psi \right] \nonumber .  
\end{eqnarray} 
Changing the order of integration and performing the Gaussian integral over
the Grassmann fields, we obtain
\begin{equation}   \label{bcs:37}
Z = \int {\rm D} \Delta^* {\rm D} \Delta \, \exp \left(i S_{\rm eff} [
\Delta^*, \Delta] + \frac{i}{\lambda_0}
\int_x |\Delta|^2 \right),  
\end{equation}
with $S_{\rm eff}$ the one-loop effective action which, using the
identity ${\rm Det}(A) = \exp[{\rm Tr} \ln(A)]$, can be rewritten as
\begin{equation}  \label{bcs:312}
S_{\rm eff}[\Delta^*, \Delta] = -i \, {\rm Tr} \ln \left(
\begin{array}{cc} p_{0} - \xi ({\bf p}) & -\Delta \\ -\Delta^* &
p_{0} + \xi ({\bf p})
\end{array}\right),
\end{equation} 
where $p_0 = i \partial_0$ and $\xi({\bf p}) = \epsilon({\bf p}) - \mu_0$,
with $\epsilon({\bf p}) = {\bf p}^2/2m$.  

The trace Tr appearing here needs some explanation.  Explicitly, it is
defined as
\begin{equation}   \label{bcs:explicit}
S_{\rm eff} = -i {\rm Tr} \, \ln \left[K(p,x) \right] = -i {\rm tr} \ln\left[
K(p,x) \delta (x - y)\bigr|_{y = x} \right],
\end{equation}
where the trace tr is the usual one over discrete indices.  We
abbreviated the matrix appearing in (\ref{bcs:312}) by $K(p,x)$ so as to
cover the entire class of actions of the form
\begin{equation} 
S = \int_x \psi^\dagger(x) K(p,x) \psi(x).
\end{equation} 
The delta function in (\ref{bcs:explicit}) arises because $K(p,x)$ is
obtained as a second functional derivative of the action
\begin{equation}  
\frac{\delta^{2} S}{\delta \psi^\dagger(x) \, \delta \psi(x)} =
K(p,x)  \,  \delta (x - y) \bigr|_{y =  x},
\end{equation} 
each of which gives a delta function.  Since the action has only one
integral $\int_x$ over spacetime, one delta function remains.  Because it is
diagonal, it may be taken out of the logarithm and the effective action
(\ref{bcs:explicit}) can be written as
\begin{eqnarray}  \label{bcs:Trexplicit}
S_{\rm eff} &=& -i {\rm tr} \, \int_x 
\ln \left[ K(p,x) \right] 
\delta (x - y) \bigr|_{y = x}  \nonumber  \\ &=&
-i {\rm tr} \, \int_x \int_k {\rm e}^{i k \cdot x} \, \ln \left[ K(p,x)
\right]  {\rm e}^{-i k \cdot x}.
\end{eqnarray}
In the last step, we used the integral representation of the
delta function:
\begin{equation}
\delta (x) = \int_k {\rm e}^{-i k \cdot x},
\end{equation}
shifted the exponential function $\exp (i k \cdot y)$ to the left, which
is justified because the derivative $p_\mu$ does not operate on it, and,
finally, set $y_\mu$ equal to $x_\mu$.  We thus see that the trace Tr in
(\ref{bcs:explicit}) stands for the trace over discrete indices as well
as the integration over spacetime and over energy and momentum.  The
integral $\int_k$ arises because the effective action
(\ref{bcs:explicit}) is a one-loop result with $k_\mu$ the loop energy
and momentum.

In the mean-field approximation, the functional integral (\ref{bcs:37}) is
approximated by the saddle point:
\begin{equation}   \label{bcs:38}
Z =  \exp \left(i S_{\rm eff}
[ \Delta^*_{\rm mf}, \Delta_{\rm mf} ]  + \frac{i}{\lambda_0} \int_x
|\Delta_{\rm mf}|^2  \right),                   
\end{equation} 
where $\Delta_{\rm mf}$ is the solution of the mean-field equation
\begin{equation}     \label{bcs:gap}
\frac{\delta S_{\rm eff} }{\delta \Delta^*(x)
} = - \frac{1}{\lambda_0} \Delta (x) . 
\end{equation}
If we assume the system to be spacetime independent, so that $\Delta_{\rm
mf}(x) = \bar{\Delta}$, Eq.\ (\ref{bcs:gap}) yields the celebrated BCS gap
equation \cite{BCS}:
\begin{eqnarray}   \label{bcs:gape} 
\frac{1}{\lambda_0} &=& - i  \int_k \frac{1}{k_{0}^{2} - E^{2}(k) + i \eta}
\nonumber \\ &=& - \frac{1}{2} \int_{\bf k} \frac{1}{E({\bf k})},
\end{eqnarray} 
where $\eta$ is an infinitesimal positive constant that is to be set to
zero at the end of the calculation, and 
\begin{equation}  \label{bcs:spec}
E({\bf k}) = \sqrt{\xi^2({\bf k}) + |\bar{\Delta}|^2}
\end{equation}  
is the spectrum of the elementary fermionic excitations. 

\subsection{BCS to BEC}
For a constant pair field, the effective action can be calculated in
closed form.  Writing
\begin{equation} 
\left(
\begin{array}{cc} k_{0} - \xi ({\bf k}) & -\bar{\Delta} \\ -\bar{\Delta}^* &
k_{0} + \xi ({\bf k}) \end{array}\right) =  \left(
\begin{array}{cc} k_{0} - \xi ({\bf k}) & 0 \\ 0 &
k_{0} + \xi ({\bf k}) \end{array}\right)  - \left(
\begin{array}{cc} 0 & \bar{\Delta} \\ \bar{\Delta}^* & 0 \end{array}\right),
\end{equation} 
and expanding the second logarithm in a Taylor series, we recognize the
form 
\begin{eqnarray}  
\lefteqn{S_{\rm eff}[\bar{\Delta}^*, \bar{\Delta}] = } \nonumber \\ &&
-i \, {\rm Tr} \ln \left(
\begin{array}{cc} k_{0} - \xi ({\bf k}) & 0 \\ 0 &
k_{0} + \xi ({\bf k}) \end{array}\right) - i \, {\rm Tr} \ln
 \left(1 - \frac{|\bar{\Delta}|^2}{k_0^2 - \xi^2({\bf k})} \right),
\end{eqnarray}  
where we ignored an irrelevant constant.  The integral over the loop
energy $k_0$ can be carried out to yield for the effective Lagrangian
\begin{equation}  \label{bcs:closed}
{\cal L}_{\rm eff} = \int_{\bf k} \left[ E({\bf k}) - \xi({\bf k})
\right].
\end{equation} 
To this one-loop result we have to add the tree term
$|\bar{\Delta}|^2/\lambda_0$.  Expanding $E({\bf k})$ in a Taylor series, we
see that the effective Lagrangian also contains a term quadratic in
$\bar{\Delta}$.  This term diverges in the ultraviolet.  To render the
theory finite to this order, we have to introduce a renormalized coupling
constant $\lambda$ defined by:
\begin{equation} \label{bcs:reng}
\frac{1}{\lambda} = \frac{1}{\lambda_0} + \frac{1}{2} \int_{\bf k}
\frac{1}{|\xi({\bf k})|}.
\end{equation} 
To this order in the loop expansion there is no renormalization of the
chemical potential, so that we can write $\mu = \mu_0$.  We regularize
the diverging integral in Eq. (\ref{bcs:reng}) by introducing a momentum
cutoff $\Lambda$.  In, for example, $d=3$, we then obtain
\begin{equation} \label{bcs:ren}
\frac{1}{\lambda} = \frac{1}{\lambda_0} + \frac{m}{2 \pi^2}
\Lambda,
\end{equation} 
where we omitted the (irrelevant) finite part of the integral.  It should be
remembered that the bare coupling constant $\lambda_0$ is negative, so that
the interaction between the fermions is attractive.  We can distinguish two
limits.  One, the famous weak-coupling BCS limit, which is obtained by
taking the bare coupling constant to zero, $\lambda_0 \rightarrow 0^-$.
Second, the limit which is obtained by letting $\lambda_0 \rightarrow -
\infty$.  In this limit, the two-particle interaction is such that the
fermions form tightly bound pairs of mass $2m$ \cite{Eagles,Leggett}.  

To explicate this so-called Bose-Einstein condensation (BEC) limit in
$d=2$, we swap the bare coupling constant for a more convenient
parameter, namely the binding energy $\epsilon_a$ of a fermion pair in
vacuum \cite{RDS}.  Both parameters characterize the strength of the
contact interaction.  To see the connection between the two, let us
consider the Schr\"odinger equation for the problem at hand.  In reduced
coordinates, it reads
\begin{equation} 
\left[- \frac{\nabla^2}{m} + \lambda_0 \, \delta({\bf x}) \right] \psi({\bf
x}) = - \epsilon_a \psi({\bf x}),
\end{equation} 
where the reduced mass is $m/2$ and the delta-function potential, with
$\lambda_0 < 0$, represents the attractive contact interaction ${\cal
L}_{\rm i}$ in the BCS Lagrangian (\ref{bcs:BCS}).  We stress that this is a
two-particle problem in vacuum; it is not the famous Cooper problem of two
interacting fermions on top of a filled Fermi sea.  The equation is most
easily solved by Fourier transforming it.  This yields the bound-state
equation
\begin{equation} 
\psi({\bf k}) = - \frac{\lambda_0}{{\bf k}^2/m + \epsilon_a} \psi(0),
\end{equation} 
or
\begin{equation} 
- \frac{1}{\lambda_0} = \int_{\bf k} \frac{1}{{\bf k}^2/m + \epsilon_a} =
\frac{1}{2} \nu(0) \ln \! \left( \frac{2 \epsilon_\Lambda}{\epsilon_a}
\right),
\end{equation} 
where $\nu(0) = m/2 \pi$ is the two-dimensional density of states (per
spin degree of freedom), and $\epsilon_\Lambda = \Lambda^2/2m$.  This
equation allows us to replace the bare coupling constant $\lambda_0$ with
the binding energy $\epsilon_a$.  When substituted in the gap equation
Eq.\ (\ref{bcs:gape}), the latter becomes
\begin{equation} \label{bcs:reggap}
\int_{\bf k} \frac{1}{{\bf k}^2/m + \epsilon_a} = \frac{1}{2}
\int_{\bf k} \frac{1}{E({\bf k})}.
\end{equation} 
By inspection, it is easily seen that this equation has a solution
\cite{Leggett}
\begin{equation} \label{comp:self}
\bar{\Delta} \rightarrow 0, \;\;\;\;\; \mu_0 \rightarrow - \tfrac{1}{2}
\epsilon_a,
\end{equation}   
with a negative chemical potential.  This is the strong-coupling BEC
limit.  To appreciate the physical significance of the specific value
found for the chemical potential in this limit, we note that the
spectrum $E_{\rm b}({\bf q})$ of the two-fermion bound state measured
relative to the pair chemical potential $2\mu_0$ reads
\begin{equation} 
E_{\rm b}({\bf q}) = - \epsilon_a + \frac{{\bf q}^2}{4m} -2 \mu_0.
\end{equation} 
The negative value for $\mu_0$ found in (\ref{comp:self}) is precisely the
condition for a Bose-Einstein condensation of the composite bosons in the
${\bf q} = 0$ state---whence the name BEC limit.

Since there are two unknowns contained in the theory, viz.\
$\bar{\Delta}$ and $\mu$, a second equation is needed to determine these
variables in the mean-field approximation \cite{Leggett}.  It is
provided by the requirement that the average fermion number $N$, which
is obtained by differentiating the effective action (\ref{bcs:312}) with
respect to $\mu$
\begin{equation} 
N = \frac{\partial S_{\rm eff}}{\partial \mu},
\end{equation} 
be fixed.  If the system is spacetime independent, this reduces to
\begin{equation} \label{bcs:n}
\bar{n} = - i\, {\rm tr} \int_k \, G(k) \tau_3,
\end{equation} 
where $\bar{n}=N/V$, with $V$ the volume of the system, is the constant
fermion number density, $\tau_3$ is the diagonal Pauli matrix in Nambu space,
\begin{equation} 
\tau_3 = \left(
\begin{array}{cr} 1 & 0 \\ 0 & -1
\end{array} \right),
\end{equation} 
and $G(k)$ is the Green function,
\begin{eqnarray}    \label{bcs:prop}
G(k) &=&
\left( \begin{array}{cc} k_0 - \xi  ({\bf k}) 
& -\bar{\Delta} \\ -\bar{\Delta}^*  & k_0 + \xi ({\bf k}) 
\end{array} \right)^{-1}  \\ &=& 
\frac{1}{k_0^2 - E^2({\bf k}) + i  \eta } 
\left( \begin{array}{cc} k_{0} \, {\rm e}^{i k_0 \eta } + \xi
({\bf k})  & 
\bar{\Delta} \\ \bar{\Delta}^* & k_{0} \, {\rm e}^{-i k_0 \eta}- \xi
({\bf k}) \end{array} \right). \nonumber 
\end{eqnarray}
Here, $\eta$ is an infinitesimal positive constant that is to be set to zero
at the end of the calculation.  The exponential functions in the diagonal
elements of the Green function are an additional convergence factor needed in
nonrelativistic theories \cite{Mattuck}.  If the integral over the loop
energy $k_0$ in the particle number equation (\ref{bcs:n}) is carried out,
it takes the familiar form
\begin{equation} \label{bcs:ne} 
\bar{n} = \int_{\bf k} \left(1 - \frac{\xi({\bf k})}{E({\bf k})} \right).
\end{equation} 
The two equations (\ref{bcs:gape}) and (\ref{bcs:n}) determine
$\bar{\Delta}$ and $\mu$.  

\subsection{Derivative Expansion \label{bcs:s5}}
When the pair field $\Delta_{\rm}$ is spacetime-dependent, the
integrals in (\ref{bcs:312}) cannot be evaluated in closed form because the
logarithm contains energy-momentum operators and spacetime-dependent
functions in a mixed order.  To disentangle the integrals we have to resort
to a derivative expansion \cite{FAF} in which the logarithm is expanded in a
Taylor series.  Each term contains powers of the energy-momentum operator
$p_\mu$ which acts on every spacetime-dependent function to its right.  All
these operators are shifted to the left by repeatedly applying the identity
\begin{equation} 
f(x) p_\mu g(x) = (p_\mu - i \tilde{\partial}_\mu) f(x) g(x),
\end{equation} 
where $f(x)$ and $g(x)$ are arbitrary functions of spacetime and the
derivative $\tilde{\partial}_\mu = (\partial_0,-\nabla)$ acts {\it only} on
the next object to the right.  One then integrates by parts, so that all the
$p_\mu$'s act to the left where only a factor $\exp(i k \cdot x)$ stands.
Ignoring total derivatives and taking into account the minus signs that
arise when integrating by parts, one sees that all occurrences of $p_\mu$
(an operator) are replaced with $k_\mu$ (an integration variable).  The
exponential function $\exp(-i k \cdot x)$ can at this stage be moved to the
left where it is annihilated by the function $\exp(i k \cdot x)$.  The
energy-momentum integration can now in principle be carried out and the
effective action be cast in the form of an integral over a local density
${\cal L}_{\rm eff}$:
\begin{equation}   
S_{\rm eff} = \int_x {\cal L}_{\rm eff}.
\end{equation} 
This is in a nutshell how the derivative expansion works \cite{FAF}.

To apply it to the BCS model and derive the Ginzburg-Landau theory we use
the following decomposition in (\ref{bcs:312}):
\begin{equation}   \label{313}
\left( \begin{array}{cc} p_{0} - \xi  ({\bf p}) & -\Delta \\
-\Delta^*  & p_{0} + \xi ({\bf p}) \end{array} \right) =  
G_0^{-1} \left[ 1 - G_0 \,
\left( \begin{array}{cc} 0 & \Delta \\
\Delta^*  & 0 \end{array} \right) \right],       
\end{equation} 
where $G_0$ is the Green function (\ref{bcs:prop}) with $\bar{\Delta}=0$.
Apart from an irrelevant constant, this leads to the expression for the
effective action
\begin{eqnarray}   \label{316}  
S_{\rm eff} &=& - i \, {\rm Tr} \, \ln \left[ 1 - G_0 \, \left(
\begin{array}{cc} 0 & \Delta \\ \Delta^* & 0 \end{array}\right)
\right] \nonumber \\ &=& i \, \mbox{Tr} \, \sum_{\ell=1}^{\infty} \,
\frac{1}{\ell}\, \left[ G_0 \, \left(\begin{array}{cc} 0 & \Delta \\
\Delta^* & 0
\end{array}\right) \right]^\ell =: \sum_{\ell=1}^{\infty}
S_{\rm eff}^{(\ell)}. 
\end{eqnarray} 
For $\ell=1$, the trace over the $2 \times 2$ matrix immediately yields zero.
In a similar fashion all terms $S_{\rm eff}^{(\ell)}$, with $\ell$ odd, give
zero.  For the quadratic term, we obtain
\begin{equation}  \label{317} 
S_{\rm eff}^{(2)} = i \, {\rm Tr} \, \frac{1}{p_{0} + \xi ({\bf p})} \,
\Delta^* \, \frac{1}{p_{0} - \xi ({\bf p})} \, \Delta,
\end{equation}
where we recall the definition of the derivative $p_\mu$ as operating on
everything that appears to the right.  Applying the derivative expansion
rules outlined above, we can cast the quadratic term in the effective
action in the form
\begin{equation}  \label{319}
S_{\rm eff}^{(2)} = i \, {\rm Tr} \, \frac{1}{k_{0} + \xi ({\bf k})} \,
\frac{1}{k_{0} - i \, \partial_{0} - \xi ({\bf k} + i \nabla)} \, \Delta^*
\Delta.
\end{equation} 
Usually, the quartic terms are included in the Ginzburg-Landau theory
without derivatives.  We then can treat $p_0 \pm \xi ({\bf p})$ as a
c-number and
\begin{equation}     \label{320}
S_{\rm eff}^{(4)} = \frac{i}{2} \, {\rm Tr} \, \frac{1}{[\, k_{0}^2 - \xi^2
({\bf k})\, ]^{2}} \, |\Delta|^4.
\end{equation} 
We may truncate the series in (\ref{316}) here, provided the pair field
$\Delta$ is small as is the case in the vicinity of the phase
transition.

To include the temperature $T$ in the theory, we adopt the
imaginary-time approach to thermal field theory \cite{Rivers,Kapusta}.
Very briefly, it can be derived from the corresponding quantum field
theory at zero temperature simply by going over to imaginary times, $t
\rightarrow -i \tau$, and substituting
\begin{equation} \label{fun:sub}
\int \frac{{\rm d} k_{0}}{2\pi}\,g(k_{0})\rightarrow  i\, \beta^{-1}
\sum_{n} \, g( i\,\omega_{n}),                             
\end{equation} 
where $g$ is an arbitrary function, while $\omega_n$ denote the Matsubara
frequencies,
\begin{equation} \label{fun:matb}
\omega_n = \left\{ \begin{array}{ll} \pi \beta^{-1} 2n, &
\;\;\;\; \mbox{(bosonic)} \\ \pi \beta^{-1}(2n + 1), & \;\;\;\; \mbox{(fermionic)}
\end{array} \right.
\end{equation} 
with $n$ an integer and $\beta= 1/T$.  With these rules, the Minkowski
action $S$ goes over into
\begin{equation} \label{fun:MtE}
S = \int_{x} {\cal L} (t,{\bf x}) \rightarrow -i \int_{0}^{\beta} {\rm d} \tau
\int_{\bf x} {\cal L} (-i\tau, {\bf x}) =: i S^{\rm E},
\end{equation}   
where the superscript E on the action at the right-hand side is to indicate
that it pertains to Euclidean rather than to Minkowski spacetime.

For the case at hand we obtain in this way the finite-temperature action:
\begin{equation}   \label{51}
S_{\rm eff}^{\rm E} = {\rm Tr}\, \left[ \frac{1}{i \omega_{n} + \xi ({\bf
k})} \, \frac{1}{i\omega_{n} + \partial_{\tau} - \xi ({\bf k} + i \nabla)}
\Delta^* \Delta + \frac{1}{2} \frac{1}{[\omega_{n}^2 + \xi^2 ({\bf
k})]^{2}} |\Delta|^4 \right] ,
\end{equation} 
where the trace at finite temperature reads explicitly
\begin{equation}
{\rm Tr} = \int_{0}^{\beta} {\rm d} \tau \int {\rm d}^dx \frac{{\rm d}^d
k}{(2\pi)^d} \, \beta^{-1} \sum_{n}.
\end{equation} 
Let us for the moment consider only the time-independent part of the
effective action (\ref{51}) and expand in gradients.  The sums over the
Matsubara frequencies are carried out with the help of the formulas
\begin{eqnarray} \label{sums}
\beta^{-1} \sum_n \frac{1}{\omega_n^2 + \xi^2} &=& \frac{1}{2}
\frac{X}{\xi} \nonumber \\ \beta^{-1} \sum_n \frac{1}{(\omega_n^2 +
\xi^2)^2} &=& -\frac{1}{4 \xi} \frac{\partial}{\partial \xi}
\left(\frac{X}{\xi}\right) \nonumber \\ \beta^{-1} \sum_n \frac{1}{i
\omega_n + \xi}\frac{1}{(i \omega_n - \xi)^2} &=& -\frac{1}{4}
\frac{\partial}{\partial \xi} \left(\frac{X}{\xi}\right) \nonumber \\
\beta^{-1} \sum_n \frac{1}{i \omega_n + \xi}\frac{1}{(i \omega_n - \xi)^3}
&=& - \frac{X}{8 \xi^3} -\frac{\beta}{16 \xi} \frac{\partial}{\partial \xi}
\left(\frac{Y}{\xi}\right) ,
\end{eqnarray} 
where $X$ and $Y$ abbreviate the functions 
\begin{equation} 
X = \tanh(\beta \xi/2), \;\;\;\;\;\; Y = 1/\cosh^2(\beta \xi/2).
\end{equation} 
As was first shown by Drechsler and Zwerger \cite{DrZw}, most of the
momentum integrals can be performed analytically for arbitrary values of the
coupling constant in two space dimensions.  In this way, they arrived at the
time-independent Ginzburg-Landau theory describing the crossover from the
weak-coupling BCS limit to the strong-coupling BEC limit:
\begin{equation}  \label{general}
{\cal L}_{\rm eff} = c \Delta^* \nabla^2 \Delta + a |\Delta|^2 -
\tfrac{1}{2} b |\Delta|^4,
\end{equation} 
with the coefficients
\begin{eqnarray}  \label{a}  
a = \frac{\nu(0)}{2} \!\!\!\! & \Biggl[ & \!\!\!\! 2 \ln\left(\frac{4 {\rm
e}^\gamma}{\pi} \right) \theta(\mu) + \ln(\beta \epsilon_a/4) + \ln(\beta
|\mu|/2) \tanh(\beta \mu/2) \nonumber \\ & & \!\!\!\! + {\rm sgn}(\mu)
\int_{\beta |\mu|/2}^\infty {\rm d} x \, \ln(x) Y(x) \Biggr],
\end{eqnarray} 
\begin{equation} \label{b}
b = \frac{\nu(0)}{4} \left[ \frac{7 \zeta(3)}{2 \pi^2} \beta^2 \theta(\mu) -
\frac{1}{\mu^2} \tanh(\beta \mu/2) + {\rm sgn}(\mu) \frac{\beta^2}{4}
\int_{\beta |\mu|/2}^\infty {\rm d} x \frac{X(x)}{x^3} \right],
\end{equation} 
and
\begin{equation} \label{c}
c = \frac{\nu(0)}{8 m} \left[ \frac{7 \zeta(3)}{2 \pi^2} \beta^2 \mu
\theta(\mu) + \frac{\beta^2}{4} |\mu| \int_{\beta |\mu|/2}^\infty {\rm d} x
\frac{X(x)}{x^3} \right].
\end{equation} 
Here, we introduced the integration variable $x = \beta \xi/2$, $\gamma =
0.577216 \cdots$ is Euler's constant, $\theta(x)$ is the Heaviside unit step
function, ${\rm sgn}(x)$ the sign function, and $\zeta(x)$ Riemann's zeta
function, with $\zeta(3)= 1.20206 \cdots$.  (The second term at the
right-hand side of Eq. (\ref{b}) differs slightly from the corresponding
term in Ref.\ \cite{DrZw}, but is consistent with a later work by Zwerger
and collaborator \cite{SZ}.)

When studying the time-dependence of the effective action (\ref{51}),
special care has to be taken with analytic continuation.  We follow S\'a de
Melo, Randeria, and Engelbrecht \cite{MRE} and first analytic continue,
using the formula
\begin{equation} \label{ana}
\beta^{-1} \sum_n \frac{1}{i \omega_n + \xi}\frac{1}{i \omega_n - i \omega_l
- \xi} = - \left[{\rm P} \frac{1}{2 \xi + q_0} - i \pi \delta(2 \xi +
q_0)\right] X,
\end{equation}
before expanding in time derivatives.  (In Ref.\ \cite{DrZw}, the
expansion was done first, leading to results which are not consistent
with those known in the BCS limit \cite{Schmid,AT}.)  In Eq.\
(\ref{ana}), ${\rm P}$ stands for the principal part, while $\omega_l$
is a bosonic Matsubara frequency, which at the right-hand side is
analytic continued to the real axes by replacing $i\omega_l$ with $q_0 +
i \eta$.  In this way, we find for the dynamic part (in
Minkowski spacetime)
\begin{equation} \label{dynamic} 
{\cal L}_{\rm dyn} = [Q'(i \partial_0) - i \pi Q''(i \partial_0)] \Delta^*
\Delta,
\end{equation} 
with
\begin{eqnarray} 
Q'(q_0) &=& - {\rm P} \int_{\bf k} \frac{q_0}{2 \xi(2 \xi + q_0)} X
\nonumber \\ Q''(q_0) &=& \tfrac{1}{2} \nu(0) \tanh(\beta q_0/4) \theta(\mu
- q_0/2).
\end{eqnarray}  
The most important result to be noted here is that the low-energy dynamics
is dissipative when the chemical potential $\mu$ is positive \cite{MRE}.
This is because the pairs can break up and decay into a continuum of
fermionic excitations.  On the other hand, for negative values of $\mu$,
where the pairs are more tightly bound, the time-dependent Ginzburg-Landau
theory describes a purely propagating pair mode.  To investigate this point
further, let us consider the two limits in detail.  

\subsection{BCS limit}
In the weak-coupling BCS limit, where the chemical potential is well
approximated by the Fermi energy $\mu = k_{\rm F}^2/2m$, with $k_{\rm F}$
the Fermi momentum, and $\epsilon_a/\epsilon_{\rm F} \rightarrow 0$, we
recover the standard result (adjusted for the reduced space dimensionality)
\cite{AT}
\begin{equation}   \label{521}
{\cal L}_{\rm eff} = \nu(0) \left\{\Delta^* \left[ \ln \!
\left(\frac{T_0}{T}\right) - \frac{\pi}{8 T_0} \partial_{0} + \frac{3
\xi_{0}^{2} }{v_{\rm F}^{2}} \left(\partial_{0}^{2} + \frac{v_{\rm F}^2}{2}
\nabla^2\right) \right] \Delta - \frac{3 \xi_{0}^{2} }{v_{\rm F}^2}
|\Delta|^4 \right\}
\end{equation} 
with $v_{\rm F}= k_{\rm F}/m$ the Fermi velocity, $\xi_0$ the BCS
correlation length
\begin{equation} 
\xi_{0}^2 =  \frac{7 \zeta(3) }{48 \pi^2} \frac{v_{\rm F}^2}{T_0^2},
\end{equation} 
and $T_0$ the BCS transition temperature \cite{DrZw}
\begin{equation} 
T_0 = \frac{{\rm e}^\gamma}{\pi} (2 \mu \epsilon_a)^{1/2}
\end{equation} 
expressed in terms of the binding energy $\epsilon_a$.  Comparing the
two terms in Eq.\ (\ref{521}) involving time derivatives, we recognize a
series expansion in powers of $q_0/T$; the terms without derivatives
constitute an expansion in powers of $|\Delta|/T$ \cite{bosepaper}.  It
therefore follows that Eq.\ (\ref{521}) represents the first terms in a
high-temperature expansion.  Since we neglected time derivatives in the
quartic and higher order terms, we are implicitly assuming that $q_0 >>
|\Delta|$ \cite{AT}.  The time-dependent Ginzburg-Landau theory
(\ref{521}) obtained in the weak-coupling BCS limit is of the
dissipative type often used in the context of defect formation in a
symmetry-breaking phase transition (for reviews see Ref.\ \cite{nato}).
For applications in the context of superconductivity see, for example,
Refs.\ \cite{DG,Tinkham,Crisan}.

\subsection{BEC Limit}
In the strong-coupling BEC limit, the general form of the effective
action (\ref{general}) and (\ref{dynamic}) reduces to \cite{DrZw}
\begin{equation}  \label{GP}
{\cal L}_{\rm eff} = \hat{\Delta}^* \left( i \partial_0 + \mu_{\rm b} +
\frac{\nabla^2}{4 m} \right) \hat{\Delta} - \lambda_{\rm b}
|\hat{\Delta}|^4,
\end{equation} 
where we introduced a rescaled pair field
\begin{equation} 
\hat{\Delta} = \left(\frac{\nu(0)}{4 |\mu|}\right)^{1/2} \Delta.
\end{equation} 
The effective theory (\ref{GP}) is precisely of the form of a
Gross-Pitaevski theory \cite{GrPi}, describing a weakly interacting
composite Bose gas with a mass $2m$ (as expected), a small chemical
potential which vanishes on approaching the critical temperature from
below [see Eq.\ (\ref{comp:self})]
\begin{equation} 
\mu_{\rm b} = 2 |\mu| \, \ln \left(\frac{\epsilon_a}{2|\mu|}\right),
\end{equation} 
or using Eq.\ (\ref{bcs:reggap}), 
\begin{equation} 
\mu_{\rm b} = \frac{|\bar{\Delta}|^2}{\epsilon_a},
\end{equation} 
and a repulsive contact interaction
\begin{equation} 
\lambda_{\rm b} = \frac{1}{\nu(0)}
\end{equation} 
independent of the binding energy $\epsilon_a$ characterizing the
interaction between the electrons.  This is special to two dimensions; in
$d$ space dimensions we have instead \cite{Pamporova}
\begin{equation} \label{comp:lambda}
\lambda_{\rm b} = (4 \pi)^{d/2} \frac{1-d/4}{\Gamma(2-d/2)}
\frac{\epsilon_a^{1-d/2}}{m^{d/2}},
\end{equation}   
with $\Gamma(x)$ the Gamma function.  Note the absence of any
temperature-dependence in the effective theory (\ref{GP}).  At zero
temperature, exactly the same effective theory was obtained by Hausmann,
using a self-consistent Green function method \cite{Haussmann}.  (See
Ref.\ \cite{Pamporova} for a derivation along the lines presented here.)
This is, as that author argued, because in the BEC limit, the critical
temperature $T_0$ is much smaller than the dissociation temperature
$T_{\rm diss} \approx \epsilon_a$ at which the tightly bound fermion
pairs are broken up by thermal fluctuations.  Hence, for all
temperatures in the range $T \leq T_0 << T_{\rm diss}$ we are
effectively in the zero-temperature regime \cite{Haussmann}.

The Gross-Pitaevski theory (\ref{GP}) describes a gapless, purely
propagating mode, viz.\ the Goldstone mode associated with the
spontaneously broken global U(1) symmetry---the so-called
Anderson-Bogoliubov mode \cite{bosepaper}.

The spacetime-dependent effective theory of a superconductor can only be
derived in a few special cases \cite{AT}: one being close to the
transition temperature where the expansion parameter is $1/T$ and
another being close to the absolute zero of temperature where the
expansion parameter is $1/|\bar{\Delta}|$.  Outside these regimes, the
effective theory depends on the ratio $|\nabla|/\partial_0$, and an
expansion in both time derivatives and gradients is not possible.  It is
amusing to note that in going from the weak-coupling BCS limit to the
strong-coupling BEC limit, we move from one valid regime to the other.

\section{Dual Theory \label{sec:dual}}
In this section we investigate the dual formulation of the
time-independent Ginzburg-Landau theory.  This approach, in which the
magnetic vortices of a superconductor play the central role, originates
from lattice studies that started more than two decades ago
\cite{BMK,Peskin,TS,HeMu,DaHa,Kleinerttri,Bartholomew,Savit}.  These
were in turn instigated by the success of the Kosterlitz-Thouless theory
describing the phase transition in a superfluid film as the unbinding of
vortex-antivortex pairs \cite{Berezinskii,KT73}.  In the dual
formulation of the Ginzburg-Landau theory, the superconductor-to-normal
phase transition is understood as a proliferation of magnetic vortices.
A detailed presentation of these matters as well as an extensive list of
references to the literature can be found in Ref.\ \cite{GFCM}.

\subsection{Electric Current Loops}
To account for the magnetic interaction we couple the Ginzburg-Landau
theory in the usual, minimal way to a vector potential ${\bf A}$.  We
also rescale the pair field such that $\beta$ times the Hamiltonian
becomes
\begin{equation}  \label{gl:H}   
{\cal H}_{\rm GL} = \left|(\nabla -  2 i e {\bf A})\phi\right|^2 +
m_\phi^2 |\phi|^2 + \lambda |\phi|^4 +
\frac{1}{2} (\nabla \times {\bf A})^2 + \frac{1}{2 \alpha} (\nabla \cdot
{\bf A})^2 ,
\end{equation}	  
where we added a gauge-fixing term with parameter $\alpha$.  To acquire a
physical understanding of what this Hamiltonian describes \cite{habil}, we
recall that a $|\phi|^4$-theory gives a field theoretic description of
strings with contact repulsion \cite{Symanzik}.  This equivalence rests on
Feynman's observation \cite{Feynman50} that the Green function
\begin{equation} \label{gl:start}
G({\bf x}) = \int_{\bf k} \frac{{\rm e}^{i {\bf k}
\cdot {\bf x}}}{{\bf k}^2 + m_\phi^2}
\end{equation}
can be expressed as a path integral.  An easy way to see this is to
invoke Schwinger's proper-time method, which is based on Euler's form
\begin{equation} \label{gl:gamma}
\frac{1}{a^z} = \frac{1}{\Gamma(z)} \int_0^\infty \frac{\dd \tau}{\tau} \,
\tau^z \, {\rm e}^{- \tau a},
\end{equation} 
to write the right-hand side of (\ref{gl:start}) as \cite{proptime}:
\begin{eqnarray} \label{gl:green}
G({\bf x}) &=& \int_0^{\infty} \dd \tau \, {\rm e}^{-\tau m_\phi^2}
\int_{\rm k} {\rm e}^{i {\bf k} \cdot {\bf x} } {\rm e}^{-\tau{\bf k}^2}
\nonumber \\ &=& \int_0^{\infty} \dd \tau \, {\rm e}^{-\tau m_\phi^2}
\left( \frac{1}{4 \pi \tau} \right)^{3/2} {\rm e}^{-\frac{1}{4} {\bf
x}^2/\tau}.
\end{eqnarray} 
The factor $(1/4 \pi \tau)^{3/2} \exp(-\frac{1}{4} {\bf x}^2/\tau)$
appearing here can be interpreted as describing a Brownian string
trajectory, showing that one endpoint of the string (located at ${\bf
x}$) has a Gaussian distribution with respect to its other endpoint
(located at the origin) \cite{PathI}.  If we imagine the string to be
composed of $N$ links, each of length $a$, then in the limit where $N
\rightarrow \infty$, $a \rightarrow 0$, the variable $\tau$
parameterizing the string stands for $\tau = N a^2/6$.  The integration
over $\tau$ in (\ref{gl:green}) indicates that the string can be
arbitrary long, but the weighing factor $\exp(-m_\phi^2 \tau)$
exponentially suppresses long ones in the normal phase where $m_\phi^2 >
0$.  To understand the physical meaning of this factor, let us return to
the discrete string model and write
\begin{equation} \label{tension}
m_\phi^2 = 6 [\sigma(a) -\sigma_{\rm cr}]/a,
\end{equation} 
so that it becomes
\begin{equation} 
{\rm e}^{-m_\phi^2 \tau} = {\rm e}^{-\sigma_{\rm eff}(a)  L},
\end{equation} 
with $\sigma_{\rm eff}(a) = \sigma(a) -\sigma_{\rm cr}$ the effective
string tension, and $L = N a$ the length of the string.  The continuum
limit is obtained by simultaneously letting $a \rightarrow 0$ and
$\sigma(a) \rightarrow \sigma_{\rm cr}$, in such a way that the
right-hand side of (\ref{tension}) tends to the finite value $m_\phi^2$.
We thus see that the factor $\exp(-m_\phi^2 \tau)$ weighs strings
according to their lengths.

Following Feynman \cite{Feynman50}, we write the right-hand side of
(\ref{gl:green}) as a path-integral, i.e., as a sum over all possible string
trajectories having one endpoint at ${\bf x}(0)=0$ and the other at ${\bf
x}(\tau) = {\bf x}$ \cite{Feynman48}:
\begin{equation} \label{gl:feynrep}
G({\bf x}) = \int_0^{\infty} \dd \tau \int_{{\bf x}(0)=0}^{{\bf
x}(\tau)={\bf x}}  \DD {\bf x}(\tau') \, {\rm e}^{-S_0},
\end{equation} 
with the (Euclidean) action 
\begin{equation} \label{gl:world}
S_0 = \int_0^\tau \dd \tau' \left[\tfrac{1}{4} \dot{\bf x}^2 (\tau') +
m_\phi^2 \right],
\end{equation} 
where $\dot{\bf x}(\tau) = \dd {\bf x}(\tau)/\dd \tau$.  

In a similar way, also the partition function of the free theory, which
written as a functional integral reads
\begin{equation} 
Z_0 = \int \DD \phi^* \DD \phi \, \exp\left(- \int_{\bf x} {\cal H}_0
\right),
\end{equation} 
can be be represented as a {\it path} integral, this time involving only
closed strings:
\begin{eqnarray} 
\ln(Z_0) &=& -\ln [ {\rm Det} ({\bf p}^2 + m_\phi^2)] = - {\rm Tr} \ln (
{\bf p}^2 + m_\phi^2)  \\
&=& \int_0^{\infty}
\frac{d\tau}{\tau} {\rm e}^{-\tau m^2_\phi} \int_{\bf k} {\rm
e}^{-\tau {\bf k}^2} = \int_0^\infty \frac{d \tau}{\tau} \oint \DD {\bf
x} (\tau') {\rm e}^{-S_0}. \nonumber
\end{eqnarray} 
Here, we used Euler's form (\ref{gl:gamma}) in the limit of small $z$.
An additional factor $1/\tau$ arises because one can start traversing a
closed string anywhere along the loop.  The $|\phi|^4$-interaction in
the Ginzburg-Landau model can be shown to result in the additional term
\cite{Parisi}
\begin{equation} 
S_\lambda  = - \lambda \int_0^{\tau} \dd \tau_l' \dd \tau_k' \, \delta \left[
{\bf x} (\tau_l') - {\bf x} (\tau_k') \right]
\end{equation} 
in the action, which gives an extra weight each time two strings---one
parameterized by $\tau_l'$ and one by $\tau_k'$---intersect.  Physically, it
represents a repulsive contact interaction between strings.  Finally, the
coupling of the field $\phi$ to the magnetic vector potential ${\bf A}$ via
the electric current 
\begin{equation} 
{\bf j}_e = - 2 e i (\phi^* \nabla \phi - \phi \nabla \phi^*) -2 (2e)^2
{\bf A} |\phi|^2
\end{equation} 
with a charge $2e$, results in the extra term \cite{GFCM}
\begin{equation} 
S_e = 2ie \int_0^{\tau} \dd \tau' \, \dot{\bf x} (\tau') \cdot {\bf A}[{\bf
x}(\tau')],
\end{equation} 
showing that the strings described by the Ginzburg-Landau theory carry an
electric current.  Pasting the pieces together, we conclude that the
partition function of the Ginzburg-Landau theory can be equivalently
represented as a grand canonical ensemble of fluctuating electric current
loops, of arbitrary length and shape \cite{Copetal}:
\begin{equation} \label{eloops}
Z = \int \DD {\bf A} \, {\rm e}^{-\frac{1}{2} \int_{\bf x} \left[(\nabla
\times {\bf A})^2 + \frac{1}{\alpha} (\nabla \cdot {\bf A})^2 \right] }
\sum_{N=0}^{\infty} \frac{1}{N!} \prod_{l=1}^N \left[ \int_0^\infty
\frac{\dd \tau_l}{\tau_l} \oint \DD {\bf x}(\tau'_l) \right] {\rm e}^{-
S_{\rm GL}},
\end{equation} 
with the action
\begin{eqnarray} \label{action}
S_{\rm GL} &=& \sum_{l=1}^N \int_0^{\tau_l} \dd \tau_l' \left\{\tfrac{1}{4}
\dot{\bf x}^2(\tau_l') + m_\phi^2 + 2ie \dot{\bf x}(\tau'_l) \cdot {\bf
A}[{\bf x}(\tau'_l)]\right\} \nonumber \\ && + \lambda \sum_{l,k=1}^N
\int_0^{\tau_l} \dd \tau_l' \int_0^{\tau_k} \dd \tau_k' \, \delta \left[
{\bf x} (\tau_l') - {\bf x} (\tau_k') \right] .
\end{eqnarray} 
On entering the superconducting phase, characterized by a sign change in
the mass term of the Ginzburg-Landau theory, the effective string
tension approaches zero as 
\begin{equation} 
\sigma_{\rm eff}(a) \propto m_\phi^2 \propto (T-T_{\rm c})^{2 \nu},
\end{equation} 
with $\nu$ the correlation length exponent, and the electric current loops
proliferate.  This proliferation is also signaled by the absolute value
$|\phi|$ of the pair field which then develops a vacuum expectation value,
identifying it as the order parameter.

\subsection{Disorder Parameter}
The dual formulation of the Ginzburg-Landau gives a similar
representation of the partition function as the path integral
(\ref{eloops}); however this time not in terms of electric current, but
of magnetic vortex loops.  To derive it, we start with a trick
\cite{fort} and introduce a (hypothetical) magnetic monopole at some
point ${\bf z}$ inside the superconductor.  A monopole is a source of
magnetic flux.  Due to the Meissner effect, the flux lines emanating
from the monopole are squeezed into a flux tube.  In this way, we
managed to create a magnetic vortex at zero external field.
Electrodynamics in the presence of a monopole was first described by
Dirac \cite{Dirac} who argued that the physical local magnetic induction
${\bf h}$ is given by the combination $\nabla \times {\bf A}({\bf x}) -
{\bf B}^{\rm P}({\bf x})$.  The subtracted plastic field
\begin{equation} \label{subtr}
B_i^{\rm P} ({\bf x}) = \Phi_0 \int_{L_{\bf z}} {\rm d} y_i \, \delta ({\bf
x} - {\bf y}),
\end{equation}
with $\Phi_0 = \pi/e$ the magnetic flux quantum, removes the field of the
so-called Dirac string running along some path $L_{\bf z}$ from the location
${\bf z}$ of the monopole to infinity.  On account of Stokes' theorem, the
plastic field satisfies the equation
\begin{equation} \label{rhom} 
\nabla \cdot {\bf B}^{\rm P} ({\bf x}) = - \rho_{\rm m} ({\bf x}),
\end{equation}  
with $\rho_{\rm m} ({\bf x}) = \Phi_0 \, \delta ({\bf x} - {\bf z})$ the
monopole density.  

We continue by writing $\phi({\bf x})$ as
\begin{equation} \label{bcs:London} 
\phi({\bf x}) = \bar{\phi} \, {\rm e}^{2i \varphi ({\bf x})},
\end{equation}   
where $\bar{\phi}$ is a constant solution of the mean-field equation.  This
approximation, where the phase of the order field is allowed to vary in
space while the modulus is kept fixed, is called the London limit.  The
Ginzburg-Landau Hamiltonian then becomes after integrating out the phase
field $\varphi$:
\begin{equation}   \label{HP}
{\cal H}_{\rm GL}^{\rm P} = \frac{1}{2} (\nabla \times {\bf A} - {\bf
B}^{\rm P})^2 + \frac{1}{2} m_A^2 A_i \left( \delta_{i j} - \frac{\partial_i
\partial_j }{\nabla^2} \right) A_j + \frac{1}{2\alpha}(\nabla \cdot {\bf
A})^2,
\end{equation} 
where  
\begin{equation} \label{photonmass}
m_A = 2e |\bar{\phi}|
\end{equation}  
is the inverse magnetic penetration depth $m_A = \lambda_{\rm L}^{-1}$.  We
gave ${\cal H}$ the superscript ${\rm P}$ to indicate the presence of the
monopole.  While the Dirac string is immaterial in the normal phase, Nambu
\cite{Nambu} argued that it acquires physical relevance in the
superconducting phase where it serves as the core of the magnetic vortex or
Abrikosov flux tube originating at the monopole.

The energy $E_V$ of this configuration is easily obtained by
substituting the solution of the field equation for the gauge field
\begin{equation}  \label{clas}
A_i ({\bf x}) = \int_{\bf y} G_{i j} ({\bf x} - {\bf y})
\left[\nabla \times {\bf B}^{\rm P} ({\bf y})\right]_j,
\end{equation}  
with $G_{i j}$ the gauge-field Green function
\begin{equation} \label{gfprop} 
G{i j} ({\bf x}) = \int_{\bf k} \left(
\frac{\delta_{i j} - k_i k_j/{\bf k}^2}{{\bf k}^2+m_A^2} + \alpha
\frac{k_i k_j}{{\bf k}^4} \right) {\rm e}^{i {\bf k} \cdot {\bf x}},
\end{equation}
back into the Hamiltonian (\ref{HP}).  The energy is divergent in the
ultraviolet because in the London limit, where the mass $|m_\phi|$ of
the $\phi$-field is taken to be infinite, the vortices are considered to
be ideal lines.  For a finite mass, a vortex core has a typical width of
the order of the coherence length $\xi =1/|m_\phi|$.  This mass
therefore provides a natural ultraviolet cutoff to the theory.  Omitting
the irrelevant (diverging) monopole self-interaction, one finds
\cite{Nambu}
\begin{equation}   \label{moncon}
E_V = \frac{1}{2} g^2 \int_{L_{\bf z}} {\rm d} x_i \int_{L_{\bf z}} {\rm d}
y_i\, G ({\bf x} - {\bf y}) 
= \sigma_V L_{\bf z} .
\end{equation}
Here,
\begin{equation}  \label{Yuka}
G ( {\bf x} ) = \int_{\bf k} \frac{ {\rm e}^{i {\bf k} \cdot {\bf
x}}}{{\bf k}^2+m^2_A} = \frac{1}{4 \pi} \frac{{\rm e}^{-m_A |{\bf x}|}}{
|{\bf x}|}
\end{equation} 
is the scalar Green function,  $g$ abbreviates the combination
\begin{equation}  \label{g}
g = \Phi_0 m_A,
\end{equation}  
while $L_{\bf z}$ denotes the (infinite) length of the Dirac string and
\begin{equation}  \label{M} 
\sigma_V= \frac{1}{8\pi} g^2 \ln\left(
\frac{|m_\phi|^2}{m_A^2} \right) = \frac{1}{4\pi} g^2 \ln (
\kappa_{\rm GL} )  
\end{equation} 
its line tension, first calculated by Abrikosov \cite{Abrikosov}.  The
value $\kappa_{\rm GL} = 1/\sqrt{2}$ of the Ginzburg-Landau parameter
\begin{equation} \label{gl:gl}
\kappa_{\rm GL}= \frac{|m_\phi|}{m_A} = \frac{\lambda_{\rm L}}{\xi},
\;\;\;\;\; \mbox{or} \;\;\;\;\; \kappa^2 _{\rm GL} =
\frac{\lambda}{2e^2},
\end{equation} 
separates the type-II regime $(\kappa_{\rm GL} > 1/\sqrt{2})$, where
isolated vortices can exist, from the type-I regime $(\kappa_{\rm GL} <
1/\sqrt{2})$, where a partial penetration of an external field is
impossible.  Remembering that $L_{\bf z}$ was the Dirac string, we see
from (\ref{moncon}) that in the superconducting phase it indeed becomes
the core of the magnetic vortex originating at the monopole, as was
first observed by Nambu \cite{Nambu}.

The operator $V(L_{\bf z})$ describing the monopole with its emerging
flux tube \cite{Marino,KRE} is easily constructed by noting that in the
functional-integral approach, a given field configuration is weighted
with a Boltzmann factor $\exp \left(-\int_{\bf x} {\cal H}_{\rm GL}^{\rm
P}\right)$, with the Hamiltonian given by Eq.\ (\ref{HP}).  From this we
infer that the explicit form of the vortex operator is
\begin{equation} \label{dop} 
V(L_{\bf z}) = \exp \left\{ \int_{\bf x} \left[(\nabla \times {\bf A})
\cdot {\bf B}^{\rm P} - \tfrac{1}{2} \left({\bf B}^{\rm P}
\right)^2  \right] \right\}.
\end{equation} 

We next demonstrate that this operator can be used to distinguish the
superconducting from the normal phase \cite{Marino,KRE}.  To this end,
let us consider the correlation function $\langle V( L_{{\bf z}} ) V^*(
L_{\bar {\bf z}}) \rangle$, where $V^*(L_{\bar {\bf z}})$ describes an
additional antimonopole brought into the system at ${\bar {\bf z}}$,
with $L_{\bar {\bf z}}$ being the accompanying Dirac string running from
infinity to ${\bar {\bf z}}$.  Since all the integrals involved are
Gaussian, this expectation value can be evaluated directly.  We proceed,
however, in an indirect way for reasons that will become clear when we
proceed, and first linearize the functional integral over the gauge
field by introducing an auxiliary field $\tilde{\bf h}$.  In the gauge
$\nabla \cdot {\bf A} = 0$, which corresponds to setting $\alpha=0$ in
the Hamiltonian (\ref{HP}), we find \cite{MA}
\begin{eqnarray}    \label{Vdu}
\lefteqn{\langle V( L_{\bf z} ) V^*(L_{\bar {\bf z}}) \rangle = }
\\ && \int {\rm D}
{\bf A} {\rm D} \tilde{\bf h} \exp \left\{ \int_{\bf x}  \left[ -\frac{1}{2}
\tilde{\bf h}^2 + i \tilde{\bf h} \cdot (\nabla \times {\bf A} -
{\bf B}^{\rm P}) - \frac{m_A^2}{2}  {\bf A}^2 \right] \right\},
\nonumber 
\end{eqnarray} 
where the plastic field satisfies (\ref{rhom}) with the monopole density
given by
\begin{equation} \label{mdens} 
\rho_{\rm m} ({\bf x}) = \Phi_0 [ \delta ({\bf x}-{\bf z})- \delta({\bf
x} - {\bar {\bf z}})].
\end{equation}   
To appreciate the physical relevance of the auxiliary field, let us
consider its field equation
\begin{equation}
\label{phys} 
\tilde{\bf h} = i (\nabla \times {\bf A} - {\bf B}^{\rm P}) = i {\bf h}.
\end{equation}
It tells us that apart from a factor $i$, the auxiliary field
$\tilde{\bf h}$ can be thought of as representing the local magnetic
induction ${\bf h}$.

The integral over the vector potential is easily carried out by
substituting the field equation for ${\bf A}$,
\begin{equation}
\label{fieldeq} 
{\bf A} = \frac{i}{m_A^2} \nabla \times \tilde{\bf h},
\end{equation} 
back into (\ref{Vdu}), with the result 
\begin{equation}    \label{Vdua}
\langle V( L_{\bf z} ) V^*(L_{\bar {\bf z}}) \rangle =
\int {\rm D} \tilde{\bf h} \, \exp \left\{ -\frac{1}{2} \int_{\bf x} \left[
\frac{1}{m_A^2}(\nabla \times \tilde{\bf h})^2 + \tilde{\bf h}^2 \right]
- i \int_{\bf x} \tilde{\bf h} \cdot {\bf B}^{\rm P} \right\}.
\end{equation} 
This shows that the magnetic vortex described by the plastic field ${\bf
B}^{\rm P}$ couples to the fluctuating massive vector field $\tilde{\bf
h}$, with a coupling constant $g$ introduced in Eq.\ (\ref{g}).  As the
temperature approaches the critical temperature from below, $\bar{\phi}$
tends to zero, so that the vortex decouples from the auxiliary field
$\tilde{\bf h}$.  The finite magnetic penetration depth in the
superconducting phase is reflected by the mass term in (\ref{Vdua}).

\begin{figure}
\begin{center}
\epsfxsize=6cm
\mbox{\epsfbox{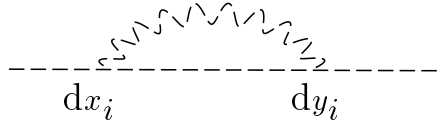}}
\end{center}
\vspace{-.5cm}
\caption{Biot-Savart interaction (wiggly line) between two line elements
${\rm d} x_i$ and ${\rm d} y_i$ of a magnetic vortex (straight line).
\label{fig:biotsavart} }
\end{figure}

After carrying out the integral over $\tilde{\bf h}$ in (\ref{Vdua}), we
obtain for the correlation function
\begin{eqnarray}  \label{vv*}
\lefteqn{\langle V( L_{\bf z} ) V^*(L_{\bar {\bf z}}) \rangle =} \\ &&
\exp \biggl\{ - \frac{1}{2} \int_{{\bf x}, {\bf y}} \bigl[
 \rho_{\rm m} ({\bf x})\, G({\bf x} - {\bf y})
\, \rho_{\rm m} ({\bf y}) + m_A^2 B^{\rm P}_i
({\bf x}) \, G({\bf x} - {\bf y}) \, B^{\rm P}_i ({\bf y}) \bigr]
\biggr\}. \nonumber
\end{eqnarray} 
The first term in the argument of the exponential function contains a
diverging monopole self-interaction for ${\bf x} = {\bf y}$.  This
divergence is irrelevant and can be eliminated by defining a
renormalized operator
\begin{equation} 
\label{renopV}
V_{\rm r} (L_{\bf z}) = V(L_{\bf z}) \exp\left[ \tfrac{1}{2} \Phi_0^2 G(0)
\right].
\end{equation} 
The second term in the argument is the most important one for our purposes.
It represents a Biot-Savart interaction between two line elements ${\rm d} x_i$
and ${\rm d} y_i$ of the magnetic vortex (see Fig.~\ref{fig:biotsavart}).
For the renormalized operators we find
\begin{equation}  \label{correlation} 
\langle V_{\rm r}( L_{\bf z} ) V_{\rm r}^*( L_{\bar {\bf z}} ) \rangle =
\exp(-\sigma_V L_{{\bf z}{\bar {\bf z}}} ) \exp\left(\frac{\Phi_0^2}{4 \pi}
\frac{ {\rm e}^{-m_A L_{{\bf z}{\bar {\bf z}}} } }{ L_{{{\bf z}}{\bar
{\bf z}}} } \right),
\end{equation} 
where $L_{{\bf z}{\bar {\bf z}}}$ denotes the length of the flux tube
connecting the monopole at ${{\bf z}}$ with the antimonopole at ${\bar
{\bf z}}$.  Initially, the two Dirac strings may run to infinity along
two arbitrary paths.  Due to the string tension, however, they join on
the shortest path $L_{{{\bf z}}{\bar {\bf z}}}$ connecting the
monopoles.

The correlation function (\ref{correlation}) behaves differently in the
two phases.  In the superconducting phase, where $m_A \neq 0$, the first
factor dominates.  It represents the confining linear potential between
the monopole and the antimonopole.  As a result, the correlation
function decays exponentially for distances larger than $1/\sigma_V$:
\begin{equation}  \label{correlator}
\left\langle V_{\rm r}( L_{z{\bar z}} ) V_{\rm r}^*(L_{z{\bar z}}) 
\right\rangle \rightarrow 0.
\end{equation}
On the other hand, in the normal phase, where the gauge field is
massless, the confinement factor in the correlation function
(\ref{correlation}) disappears, while the argument of the second
exponential turns into a pure Coulomb potential.  The correlation
function remains, consequently, finite for large distances:
\begin{equation}
\left\langle V_{\rm r}( L_{z{\bar z}} ) V_{\rm r}^*(L_{z{\bar z}}) 
\right\rangle \rightarrow 1,
\end{equation} 
thus indicating a proliferation of magnetic vortices in the normal
phase.  This demonstrates that the vortex operator can be used as an
order parameter to distinguish the two phases.  Since it develops a
vacuum expectation value not in the superconducting, but in the normal
phase it is referred to as a {\it disorder} parameter.

It is interesting to consider in detail the limit $T \uparrow T_{\rm
c}$, where the coupling constant $g = \Phi_0 m_A$ tends to zero and the
magnetic vortex decouples from the massive vector field \cite{MA}.  In
Eq.\ (\ref{Vdua}), this limit yields the constraint $\nabla \times
\tilde{\bf h} = 0$ which can be solved by setting $\tilde{\bf h} =
\nabla \gamma$.  The correlation function then takes the simple form
\begin{equation} \label{dual:simple} 
\langle V_{\rm r}({\bf z}) V_{\rm r}^*( {\bar {\bf z}}) \rangle = \int
{\rm D} \gamma \exp \left[ -\frac{1}{2} \int_{\bf x} (\nabla \gamma)^2 +
i \int_{\bf x} \gamma \rho_{\rm m} \right].
\end{equation} 
In the absence of monopoles, the theory reduces to that of a free
gapless mode $\gamma$ that may be thought of as representing the
magnetic scalar potential:
\begin{equation} 
\label{gammaid}
\nabla \gamma = i \nabla \times {\bf A} .
\end{equation} 
This follows from combining the physical interpretation of the vector
field ${\bf h}$ (\ref{phys}) with the equation $\tilde{\bf h} = \nabla
\gamma$.

The correlation function (\ref{dual:simple}) can be put in the form
\begin{equation}    \label{localrep}
\langle V_{\rm r}({\bf z} ) V_{\rm r}^*(
{\bar {\bf z}}) = \left\langle {\rm e}^{i \Phi_0 [\gamma ({\bf z})-
\gamma({\bar {\bf z}})] } \right\rangle,
\end{equation}  
where the average at the right-hand side is taken with respect to the
free scalar theory.  This equation shows that in the normal phase ($T >
T_{\rm c}$), the Dirac string looses its physical relevance, the
right-hand side depending only on the end points ${\bf z}$ and ${\bar
{\bf z}}$, not on the path $L_{{\bf z}{\bar {\bf z}}}$ connecting these
points.  The notion of a magnetic vortex is of no relevance in this
phase because the vortices proliferate and carry no energy.  This is the
reason for omitting any reference to vortex lines in the argument of $V$
in Eqs.\ (\ref{dual:simple}) and (\ref{localrep}).

\subsection{Magnetic Vortex Loops}
From the above results for a single vortex, we can now easily infer the
dual formulation of the Ginzburg-Landau theory.  In this formulation,
the partition function is written as a grand canonical ensemble of
fluctuating magnetic vortex loops, of arbitrary length and shape:
\begin{equation}  \label{pathZ} 
Z = \int \DD \tilde{\bf h} \, {\rm e}^{-\frac{1}{2} \int_{\bf x} \left[
 (\nabla \times \tilde{\bf h})^2/m_A^2 + \frac{1}{2} \tilde{\bf
h}^2 \right]} \sum_{N=0}^{\infty} \frac{1}{N!} \prod_{l=1}^N \left[
\int_0^\infty \frac{\dd \tau_l}{\tau_l} \oint \DD {\bf x}(\tau'_l) \right]
{\rm e}^{- S_{\rm dual}}
\end{equation}  
with the dual action, [cf.\ Eq.\ (\ref{action})]
\begin{eqnarray} 
S_{\rm dual} &=& \sum_{l=1}^N \int_0^{\tau_l} \dd \tau_l' \left\{\tfrac{1}{4}
\dot{\bf x}^2(\tau_l') + m_\psi^2 + i \Phi_0 \dot{\bf x}(\tau'_l) \cdot
\tilde{\bf h}[{\bf x}(\tau'_l)]\right\} \nonumber \\ && + u
\sum_{l,k=1}^N \int_0^{\tau_l} \dd \tau_l' \int_0^{\tau_k} \dd \tau_k'
\, \delta \left[ {\bf x} (\tau_l') - {\bf x} (\tau_k') \right] .
\end{eqnarray} 
We have included here a mass term $m^2_\psi$ representing the intrinsic
vortex line tension [cf.\ Eq.\ (\ref{tension})], and also a contact
repulsion between the vortices parameterized by $u$.

The equivalent field representation of this dual theory reads
\cite{BS,Kaw,GFCM,KKR,MA,fort},
\begin{equation}   \label{funcZ}
Z = \int \DD \tilde{\bf h} \DD \psi^{*} \DD \psi 
\,  \exp \left(- \int_{\bf x} {\cal H}_{\rm dual}\right)
\end{equation}  
with the Hamiltonian
\begin{equation}     \label{Hpsi}
{\cal H}_{\rm dual} =   \frac{1}{2 m_A^2} (\nabla \times 
\tilde{\bf h})^2 + \frac{1}{2} \tilde{\bf h}^2 + |(\nabla -i \Phi_0
\tilde{\bf h}) \psi|^2 + m_\psi^2 |\psi|^2 + u |\psi|^4 ,   
\end{equation} 
where the disorder field $\psi$ is minimally coupled to the massive
vector field $\tilde{\bf h}$, representing the local magnetic induction.
The dual formulation contains the same information as the original,
Ginzburg-Landau formulation.  For example, the vortex line tension
(\ref{M}) appears in the dual theory as a one-loop on-shell mass
correction stemming from the graph depicted in Fig.\
\ref{fig:biotsavart} which we now interpret as a Feynman graph of the
dual theory (\ref{Hpsi}), with the straight and wiggly line denoting
respectively the $\psi$- and $\tilde{\bf h}$-field Green function.  Also
the fixed points of the Ginzburg-Landau theory map onto those of the
dual field theory \cite{KKR}.  And both formulations can be used to
study the critical behavior of the superconductor-to-normal phase
transition (see Refs.\
\cite{Peskin,Kleinerttri,GFCM,PRL,fort,Herbut,CaNo} for the dual
approach).
\\ 

\noindent
{\bf Acknowledgments} 

\noindent
It is a pleasure to thank N. Antunes, L. Bettencourt, A. Leggett, D. Steer,
and G. Volovik for useful discussions during the NATO Winter School and
European Science Foundation (ESF) Workshop {\it Topological Defects and the
Non-Equilibrium Dynamics of Symmetry Breaking Phase Transitions}, Les
Houches, February 16-26, 1999, and M. Crisan for helpful correspondence.

This work is performed as part of a scientific network supported by the
ESF (see network's URL, http://www.physik.fu-berlin.de/$\sim$defect).

\end{document}